\documentclass[aps,twocolumn,amsmath,amssymb]{revtex4}

\usepackage{graphicx}

\renewcommand{\>}{\rangle}
\newcommand{\<}{\langle}
\newcommand{\e}{\rm e}

\begin{document}

\title{Effect of frequency mismatched photons in quantum information processing}

\author{J. \ Metz} \email{jeremy.metz@imperial.ac.uk}
\affiliation{Blackett Laboratory, Imperial College London, Prince
Consort Road, London SW7 2BW, United Kingdom}

\author{S. D.\ Barrett} \email{seandbarrett@gmail.com}
\affiliation{Blackett Laboratory, Imperial College London, Prince
Consort Road, London SW7 2BW, United Kingdom}

\date{\today}

\begin{abstract}
Many promising schemes for quantum information processing
(QIP) rely on few-photon interference effects. In these proposals,
the photons are treated as being indistinguishable particles.
However, single photon sources are typically subject to variation
from device to device. Thus the photons emitted from different
sources will not be perfectly identical, and there will be some
variation in their frequencies. Here, we analyse the effect of this
frequency mismatch on QIP schemes. As examples, we consider the
distributed QIP protocol proposed by Barrett and Kok
\cite{BarrettKok2005}, and Hong-Ou-Mandel interference which lies at
the heart of many linear optical schemes for quantum computing
\cite{KLM,KokReview,Legero2003,Legero2004}. In the distributed QIP
protocol, we find that the fidelity of entangled qubit states
depends crucially on the time resolution of single photon detectors.
In particular,
There is no reduction in the fidelity when an ideal detector model
is assumed, while reduced fidelities may be encountered when using
realistic detectors with a finite response time. We obtain similar
results in the case of Hong-Ou-Mandel interference -- with perfect
detectors, a modified version of quantum interference is seen, and
the visibility of the interference pattern is reduced as the
detector time resolution is reduced. Our findings indicate that
problems due to frequency mismatch can be overcome, provided sufficiently
fast detectors are available.
\end{abstract}

\maketitle

\section{Introduction}

Few-photon interference effects are of fundamental interest, as they
have no classical analogue, and demonstrate the quantum nature of
the radiation field. The archetypal example of such an effect is
two-photon interference as observed by Hong, Ou and Mandel in
1987 \cite{HOM}. There two identical photons, each incident on
a separate input port of a beam splitter, coalesce such that both
photons are \emph{always} detected at the same output mode of the beam
splitter. More recently, single-photon interference effects have
been proposed as a means to entangle remote pairs of atomic systems
\cite{Cabrillo,Bose}. Here, the atoms emit photons which are then
incident on a beam splitter, followed by measurements on the output ports of the
beam splitter. The role of the beam splitter is to
coherently erase {\em `which path'} information. Since the photons are
identical and the observer cannot know which atom the photon was
emitted from, the result is to prepare the pair of atoms in an
entangled state.

Such interference effects are currently the subject of much
interest, as they potentially provide a route to scalable quantum
information processing. Few-photon interference lies at the heart of
schemes for linear optics quantum computing
\cite{KLM,YoranReznik,Nielsen,BrowneRudolph}. This is also central
to many hybrid light-matter quantum computing schemes
\cite{BarrettKok2005,LimBeigeKwek,LBBKK2006,Duan}, in which remote
matter qubit systems (such as trapped atoms, quantum dots, or
impurity centers in solids) can be entangled via single photon
interference effects, in a way such that efficient quantum
computation is possible. This approach can significantly simplify
scaling the computer to a large number of qubits. As a result there
is now growing interest from experimental groups in implementing
distributed schemes \cite{Moehring2007,Grangier}. Other applications
of single photon interference in QIP have also been proposed, such
as quantum repeaters \cite{DCLZ,Chen06i}, and are currently being
actively pursued by experimental groups \cite{KimbleExperiments}.

Few-photon interference effects are often said to require
\emph{identical} photons, such that the frequency,
polarization, and temporal envelope of each photon should be
indistinguishable.
This is because these experiments make use of a
beam splitter to erase `which path' information, so that the source
of each photon cannot be inferred from the detector signal, even in
principle.
Any additional information carried by the photon (such as
the frequency) could, in principle, be used to infer the path that
the photon took, and therefore will tend to degrade the interference.
From this perspective, one expects that few-photon
interference cannot be observed between photons from sources of different
frequency. Indeed, if one restricts ones attention to the total
coincidence rates, this is indeed what is observed in a Hong-Ou-Mandel
type experiments. For sufficiently detuned single photon sources, the
photons behave as independent particles, each exiting either port of
the beam splitter with probability ${\textstyle \frac{1}{2}}$, and no interference
is observed. However, this begs the question, {\em ``where does the
interference `go'?''} Usually we only expect quantum effects to vanish
in the presence of some kind of noise or decoherence process.

Some insight into this issue has been provided in a series of
intriguing theoretical and experimental results by Legero and
co-workers, concerning the Hong-Ou-Mandel effect with different
frequency photons \cite{Legero2003,Legero2004}. They showed
analytically that if one can perform \emph{time resolved}
measurements in a Hong-Ou-Mandel type experiment with detuned
photons, a type of quantum interference can still be observed. The
probability of both photons appearing at the same output port of the
beam splitter is now no longer a constant value, but rather
oscillates as a function of the time between the photon detection
events. The frequency of this oscillation is given by the detuning
between the two photons, and thus has been called a `quantum beat'
of two photons \cite{Legero2004}. This effect was observed
experimentally using successive photons from a single photon source
implemented by a Raman transition in an atom-cavity system
\cite{Legero2004}.

Aside from the conceptual interest in these effects, the issue of
interference between photons from non-identical sources is now of
significant practical importance, since there is much interest in
using such effects in quantum information processing. Scaling these
proposals will require observing quantum interference between
photons from many different sources. These sources may be
manufactured devices, such as quantum dots or other systems coupled
to micro-cavities \cite{Vahala2003}, and as such will be subject to
fabrication imperfections. In particular, some variation of the
relevant optical frequencies of the devices is to be expected. This
will also be a problem in certain `natural' systems, such as
nitrogen-vacancy centers in diamond, which typically experience a
spread in their resonance frequencies due to local strain fields
\cite{DiamondStrain}. While it could be possible to tune such
systems over a certain frequency range, it may nevertheless be
difficult to bring all of the sources into resonance with each
other. Thus it is important to understand the extent to which mutual
detuning of the sources induces errors in QIP schemes making use of
single photon interference effects, and what can be done to mitigate
such errors.

In this paper we investigate the practical and fundamental aspects
of the effect of detuning on few-photon interference by considering
two particular cases. Firstly, we consider the entangling
operation introduced by Barrett and Kok \cite{BarrettKok2005}. This
operation allows the preparation of entangled states of remote
qubits, and furthermore can be used to generate \emph{graph states}
of multiple qubits, and hence is a resource for scalable, universal
quantum computation. In addition we consider the Hong-Ou-Mandel effect.
This is of interest because it is one of the best known few photon
interference effects, and is also central to multiple schemes for linear
optics quantum computation \cite{KLM,KokReview} and quantum repeaters
\cite{Chen06i}. In both cases, we first consider the corresponding
effects with detuned photons in the case of idealized (i.e.~perfect
time resolution) detectors, and find that a modified version of the
entanglement/interference effect persists. We then consider the
opposite case, where the detectors have very bad time resolution,
and find that in this limit, the interference (or entanglement) is
indeed reduced. Loosely speaking, the degree of reduction of
entanglement/interference depends on the extent to which the photons
are distinguishable in the frequency domain.

By making use of an explicit model of the photodetectors
\cite{wiseman,wiseman2}, we also consider the intermediate case,
where some time resolution present in the detector outputs, and
quantify how the entanglement/interference is modified as the
detector resolution varies. Our results are of direct practical
benefit for the implementation of these schemes, as they allow one
to determine the level of error that can be expected for a given
detuning and detector resolution. Furthermore, we hope the results
will aid in understanding the nature of single photon interference
effects with detuned photons, and give some insight into where the
entanglement/interference `goes'.

Although we focus on two particular examples of few-photon
interference in this paper, the techniques are reasonably generic
and can therefore also be applied to many other schemes which
involve similar effects. The effect of frequency mismatched photons
has also been examined in slightly different contexts, such as
sources of entangled photon pairs \cite{Stace2003}. We also note
that a potential solution to the frequency mismatch problem has been
proposed \cite{Stace2006}. This scheme makes use of acousto-optic
modulators as `frequency beam splitters' which can be used to erase
the frequency information of the photons.

The remainder of this paper is organized as follows. In Section
\ref{sec:barrett+kok} we review the entangling scheme of Barrett and
Kok \cite{BarrettKok2005}. We examine the effect of photon detuning
in this scheme
 in Section \ref{sec:entangle_mismatch}. The case of ideal time resolution detectors is considered in Section \ref{sec:ideal_ent}, while in \ref{sec:bad_ent} we examine the opposite limit of very bad time resolution detectors, and in Section \ref{sec:moderate_ent} the intermediate case is analysed.
 In Section \ref{sec:HOM} we explore the influence of detuning on the Hong-Ou-Mandel effect, again in the regimes of good, bad, and intermediate time resolution detectors.
Finally, we summarize our findings and draw some conclusions in Section \ref{sec:conc}.

\section{Entangling atoms}
\label{sec:barrett+kok}

In this section, we review the method proposed in Ref.
\cite{BarrettKok2005} for entangling remote pairs of qubit systems,
which could be trapped atoms, ions, quantum dots, or impurity
centers in solids. This method actually implements a
non-deterministic parity measurement such that, when successful, a
projection of the form $\vert 01 \rangle\langle 01\vert + \vert 10
\rangle\langle 01\vert $ is performed on the joint state of the
qubits. A positive outcome is heralded by a particular sequence of
detector clicks. If these are not observed, the operation has
failed, and the qubits can be reset and the operation reattempted.
This operation, combined with single qubit rotations and
measurements, is sufficient to efficiently generate arbitrary graph
states of multiple qubits, which in turn permit universal
measurement-based quantum computation. A number of other schemes
have also been proposed for remote entanglement via few-photon
interference
\cite{Cabrillo,Bose,LimBeigeKwek,LBBKK2006,Duan,Brown2003,Feng2003,
DuanKimble2003,Simon2003,Protsenko2002,Zou2004,Engel2006}, and
although we do not consider these schemes explicitly, the results in
this paper are expected to also be applicable to those schemes.

\begin{figure}
\begin{center}
\resizebox{\columnwidth}{!}{\includegraphics {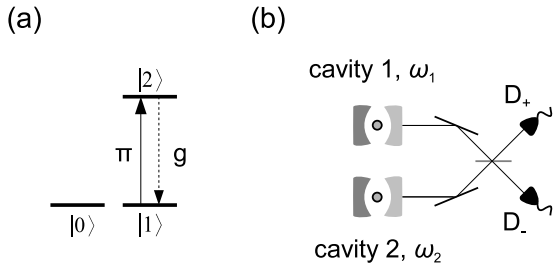}}
\end{center}
\vspace*{-0.5cm} \caption{Schematic diagram of the setup proposed.
(a) Level structure of the atomic system. $|0\rangle$ and
$|1\rangle$ are low lying, long lived states representing the qubit
degree of freedom. $|1\rangle$ is connected to $|2\rangle$ by an
optical transition which may be addressed by a $\pi$-pulse of a laser to swap population between these states, and coupled to the cavity mode with coupling constant $g$. 
(b) Setup for remote entanglement. The light
from each atom is collected (possibly with the aid of an optical
cavity on resonance with the $1 \to 2$ transition) and mixed on a
50:50 beam splitter. Photon counting detectors $D_+$ and $D_-$
monitor the output modes of the beam splitter. In the original
proposal, the frequencies $\omega_1$ and $\omega_2$ of the optical
transitions are assumed to be equal \cite{BarrettKok2005}. In
general, this will not be the case and we examine this scenario in
Section \ref{sec:entangle_mismatch}.} \label{fig:setup}
\end{figure}

The setup proposed consists of two atoms \footnote{Note that, in
general, the scheme can be applied to a variety of different qubit
realizations such as trapped atoms, ions, quantum dots, or impurity
centers in solids, provided they have the appropriate level
structure of Figure \ref{fig:setup} (a). Hereafter, we shall just
refer to the systems as `atoms' for brevity.} inside separate
cavities, as shown in Figure \ref{fig:setup}, which are assumed to have equal
resonant frequencies $\omega_1 = \omega_2$. The qubit levels $|0\>$
and $|1\>$ are long lived, low-lying states of the atoms. In
addition there is an excited level, $|2\>$, such that an optical
transition between $|1\>$ and $|2\>$ couples resonantly to the
cavity mode of the respective cavity. The use of cavities is not strictly
necessary for the ideas presented here, since the same protocol will
work if the light emitted from the optical transition is monitored
in free space. However, coupling via a cavity
may offer an increase in the success rate of the protocol with
respect to the same setup without cavities. The protocol for entangled pair
generation proceeds as follows:
\begin{enumerate}
    \item Prepare atoms in the product state $|+\> \otimes |+\>$,
    \item apply a $\pi$-pulse on the $2 \to 1$ transition to prepare the
    atoms in ${\textstyle \frac{1}{2}}(|00\>+|02\>+|20\>+|22\>)$,
    \item monitor cavity output for a time significantly longer than the decay
    time of the $2 \to 1$ transition; a click in either detector signals
    a successful first round. The absence of a click implies a
    failure of the operation, and the protocol should start again
    from step (1).
    \item Apply a bit-flip, $\sigma_x$, on the qubit states to perform $|0\> \leftrightarrow |1\>$
    on each atom.
    \item Repeat steps (2) and (3); a second click in either detector signals the
    successful completion of the protocol.
\end{enumerate}

We now review these steps in more detail. The atoms are initially
individually prepared in the $|+\> \equiv {\textstyle
\frac{1}{\sqrt{2}}}(|0\> + |1\>)$ state, such that the combined
state of the atoms may be written as
\begin{eqnarray}
\label{eqn:initialstate_bk}
    |\psi(0)\> &=& |+\> \otimes |+\> \equiv \frac{1}{2}(|00\> + |01\> + |10\> + |11\>)\,.~~~
\end{eqnarray}
Then a $\pi$-pulse is applied to excite the 1--2 transition such
that the state of the atoms becomes $\frac{1}{2}(|00\> + |02\> +
|20\> + |22\>)$. Next the cavity leakage is monitored using the
detectors $D_{+}$ and $D_{-}$. A click in either detector signals a
successful first round of the protocol. If the detection process
were perfect it would be possible to stop the protocol here and be
confident of having correctly performed an entangling operation
on the atoms. The final state of the atoms in this case is given by
\begin{eqnarray}
\label{eqn:psi1click_bk}
    |\psi^{\pm}\> &=& \frac{1}{\sqrt{2}}(|01\> \pm |10\>) \,,
\end{eqnarray}
where the sign between the terms is determined by the detector where
the click was observed. This state is obtained because the only
parts of the initial state that create {\em exactly} one
cavity excitation are initial atomic states $|01\>$ and $|10\>$. The
presence of the beam splitter erases the which-path information such
that a click in either detector will not reveal any information
about which atom/cavity the excitation originated from.

However, in general the photon collection process as well as the detectors themselves will not be perfect. Therefore the detection of a single click will lead to a mixed state over the one- and two-excitation parts of the atomic states,
\begin{eqnarray}
\label{eqn:rho1click_bk}
    \rho^{\pm} &=& p_1 |\psi^{\pm}\>\<\psi^{\pm}| + (1-p_1)|11\>\<11| \,,
\end{eqnarray}
where $p_1$ is the probability of there having been only one photon.
Note that strictly speaking this neglects the presence of dark
counts in the detectors, which would lead to a
$|00\rangle\langle00|$ contribution to the state. However, for
existing detectors the dark count rate is typically much smaller
than the atom/cavity emission rates involved in the protocol,
thereby justifying this approximation. In addition, detector
dead-times do not affect this protocol, since for a successful
outcome, only a single click is observed on each round, and
two-excitation events are rejected as a result of the post selection
process (as we describe below).

The solution proposed to overcome the presence of the two-excitation
component in the state, was to apply a bit-flip pulse to each qubit,
i.e. a $\sigma_x$ operation on $|0\>$ and $|1\>$, followed by a
second round of the protocol. The bit-flip operation has no effect
on $|\psi^{\pm}\>$, but changes $|11\> \to |00\>$. Therefore a
second round of the protocol resulting in a second click eliminates
the $|00\>$ part of the state as $|00\>$ does not couple to the relevant optical fields.
Thus no photon can result from
this component of the state. Then the final state of the system,
conditional on observing a photon in each round, is given by
\begin{eqnarray}
\label{eqn:psi2click_bk}
    |\psi\> &=& \frac{1}{\sqrt{2}} (|01\> + (-1)^m |10\>) \,,
\end{eqnarray}
where $m=0$ if both clicks occur in the same detector, or $m=1$ if
they are in different detectors. This state is maximally entangled
and independent of click times. The independence on click times is
due to the assumption that $\omega_1 = \omega_2$. We shall consider
the case of detuned cavities in the following section.

The success rate of the protocol is
\begin{eqnarray}
\label{eqn:success_bk}
    P_{\rm succ} &=& \frac{1}{2} \eta^2  \,,
\end{eqnarray}
where $\eta$ is the combined efficiency of collection and detection
of the photons, while the factor ${\textstyle \frac{1}{2}}$ is from the population of the
initial state in the $\{|01\>, |10\>\}$ subspace. In spite of the
inherent non-determinism of this protocol, efficient quantum
computing is still possible using this operation, in principle with
any success probability larger than zero \cite{BarrettKok2005}. (The
price to be paid for this is in an overhead cost for building the
cluster states, which may become impractical for very small success
probabilities \cite{RohdeBarrett2007}).

Note that the fidelity of the entangling operation is not affected
by photon collection or detection efficiency, since only outcomes in
which a photon was observed on each round of the protocol are
retained (i.e.~post-selected) as successful outcomes. This is a
useful fact for analysing the fidelity of the scheme in the presence
of other imperfections, since it means that one can essentially
ignore the detection and collection inefficiencies
in calculations, and still arrive at reliable
values for the fidelity. One should still be mindful, however, that
photon loss will lower the success probability. This is an
approach that we will adopt in the remainder of this paper.

\section{Entangling operations with detuned sources}
\label{sec:entangle_mismatch}

In this section we examine the effect of cavity frequency mismatch
on entangling operations. In particular we examine the scheme
outlined in Section \ref{sec:barrett+kok}. The setup is generalized
to cavities with different frequencies, $\omega_1$ and $\omega_2$.
The atomic transition frequencies are each still assumed to be
resonant with the corresponding cavity transition, such that
$\omega_{12;1} = \omega_{1}$ and $\omega_{12;2} = \omega_{2}$, where
$\omega_{12;j}$ is the frequency of the $1 \to 2$ transition in atom
$j$. We make this assumption partly to simplify the analysis so as
to concentrate specifically on the effect of frequency mismatch
\emph{between} two sources, but it is also reasonably well motivated
physically. This is because in systems where the atomic transition
frequencies are not naturally on resonance with the cavity
resonance, it may still be possible to tune the transitions into
resonance, for example through the use of Stark or Zeeman shifts of
the atomic levels. However, tuning the cavities so that they are
also mutually on resonance may be more difficult, especially in the
case of monolithic micro-cavities. We also restrict our attention to
the case where both cavities have the same decay rate $\kappa_1 =
\kappa_2 = \kappa$, and the same atom-cavity coupling strength,
$g_1 = g_2 = g$. A treatment when this is not the case has already
been presented in \cite{BarrettKok2005}.

Using the Quantum Jump (QJ) formalism
\cite{Carmichael1993Book,Breuer2002Book,Beige1997THS,Hegerfeldt1996QSO,Plenio1998RMP}
which is particularly well suited to describing systems under
continuous observation, we find that the Hamiltonian describing the
evolution of the system condition on no-photon emissions is given by
\begin{eqnarray}
\label{eqn:Hcond_initial}
    H_{\rm cond}^S &=& \sum_{j=1}^2 \hbar \omega_{j} (|2\>_{jj}\<2| + b_j^{\dag}b_j)
    \nonumber\\
    && + \hbar g( b_j |2\>_{jj}\<1| +  b_j^{\dag} |1\>_{jj}\<2|)
        - \frac{i \hbar \kappa}{2} b_j^{\dag}b_j  \, , ~~~
\end{eqnarray}
where the index $j=1,2$ labels the respective atom-cavity systems.
We have defined the energy of the degenerate ground states $|0\>_j,
|1\>_j$ to be zero. The first term corresponds to the energies of
the atoms and the cavity fields respectively. The second term
describes the Jaynes-Cummings interaction between the cavities and
the 1--2 transitions of the atoms, while the last term comes from
the QJ description of the cavity-free field interaction. This term
is non-Hermitian and leads to a decrease in the norm of the state
vector which quantifies the decrease in probability of the system
not emitting photons. We now transform to an interaction picture via
the unitary evolution operator $U_0 = \exp[-{\textstyle
\frac{i}{\hbar}} \omega_1 (\sum_{j=1}^2 |2\>_{jj}\<2| + b_j^{\dag}
b_j) t]$. The state transforms according to $|\psi_I(t)\> \equiv U_0^{\dag}
|\psi_S(t)\>$ where $|\psi_S (t)\>$ is the state in the
Schr\"odinger picture. The dynamics in the interaction picture is
given by $H_{\rm cond}$, where $H_{\rm cond} \equiv U_0^{\dag} H_{\rm
cond}^S U_{0} + i \hbar \dot{U}_0^{\dag} U_0$. Applying this
transformation to (\ref{eqn:Hcond_initial}) gives
\begin{eqnarray}
\label{eqn:Hcondint}
    H_{\rm cond} &=& \sum_{j=1}^2 \hbar (g_j b_j |2\>_{jj}\<1| + g_j^* b_j^{\dag} |1\>_{jj}\<2|) \nonumber\\
        && + \hbar \Delta(|2\>_{22}\<2| + b_2^{\dag}b_2)
        - i {\textstyle \frac{\hbar }{2}} \kappa b_j^{\dag}b_j \,, ~~
\end{eqnarray}
where we have defined $\Delta \equiv \omega_2 - \omega_1$. The
associated jump operators which describe the evolution of the system
in the event of an emission of a photon out of either cavity are given
by
\begin{eqnarray}
\label{eqn:jump}
    R_{1} &=& \sqrt{\kappa} b_1 \, ,\nonumber\\
    R_{2} &=& \sqrt{\kappa} b_2 \, .
\end{eqnarray}
Now the {\em unconditional} master equation for this dissipative system may be written as
\begin{eqnarray}
\label{eqn:master}
    \dot{\rho} &=& \textstyle{\frac{i}{\hbar}}(H_{\rm cond} \rho - \rho H_{\rm cond}^{\dag}) + R_1  \rho R_1^{\dag} + R_2 \rho R_2^{\dag} \,. ~~
\end{eqnarray}
The effect of a 50-50 beam splitter which mixes the cavity outputs
$b_1$ and $b_2$, as described in Section \ref{sec:barrett+kok} is
described by the beam splitter transformation,
\begin{eqnarray}
\label{eqn:beamsplitter}
    c_+ &=& \frac{b_1 + b_2}{\sqrt{2}} \, ,\nonumber\\
    c_- &=& \frac{b_1 - b_2}{\sqrt{2}} \, ,
\end{eqnarray}
where $c_+$ and $c_-$ are the two output modes of the beam splitter.
Note that the transformed operators, $c_+$ and $c_-$, do not have an
explicit time dependence in the interaction picture. This is a
consequence of the form of $U_0$ -- both cavity jump operators, $b_{1,2}$,
receive the same time-dependent phase shift, and so there is no
relative phase between the $b_1$ and $b_2$ terms in the
transformation. This transformation leaves the master equation
unchanged, but will however influence single trajectories. This is
reflected by the change in jump operators,
\begin{eqnarray}
\label{eqn:jump2}
    R_+ &=& \sqrt{\frac {\kappa}{2}} (b_1 + b_2) = \sqrt{\kappa} c_+  \,, \nonumber\\
    R_- &=& \sqrt{\frac {\kappa}{2}} (b_1 - b_2) = \sqrt{\kappa} c_- \,. ~~
\end{eqnarray}
Now we use the fact that for the over damped (i.e. Purcell) regime,
i.e. when $\kappa \gg g$, we may eliminate the populated cavity mode
to simplify the analysis \cite{Metz2007PRA}. This approximation is
possible as the population in the cavity mode remains negligible in
this regime. Then we find that the effective Hamiltonian of the
system is then given by
\begin{eqnarray}
\label{eqn:Heff}
    H_{\rm cond} &=& \hbar \Delta |2\>_{22}\<2|
    - \sum_{j=1}^2 i \textstyle{\frac{\hbar}{2}}
    \kappa_{\rm eff} |2\>_{jj}\<2| \, ,
\end{eqnarray}
where $\kappa_{\rm eff} \equiv 4 g^2 / \kappa$ is the effective
decay rate of the atoms, for decay via the cavity mode. Similarly
the effective jump operators are given by
\begin{eqnarray}
\label{eqn:jumpeff}
    R_+ &=& \sqrt{\frac{\kappa_{\rm eff}}{2}} \left(|1\>_{11}\<2| + |1\>_{22}\<2| \right)  \, ,\nonumber\\
    R_- &=& \sqrt{\frac{\kappa_{\rm eff}}{2}} \left(|1\>_{11}\<2| - |1\>_{22}\<2| \right) \, .
\end{eqnarray}
Note that Eqs.~(\ref{eqn:Heff}--\ref{eqn:jumpeff}) are also
applicable to setups with no cavities, where the spontaneously
emitted light is directly collected via a system of lenses and other
optical elements \cite{Moehring2007,Grangier}. In this case,
$\kappa_{\rm eff}$ should be replaced by the appropriate rate for
spontaneous emission into the collected mode.

\subsection{Ideal detector case}
\label{sec:ideal_ent}

In this section we consider the case of remote entangling operations
in the case where the time resolution of the detectors, $t_r$, is
`ideal', in the sense that it is very much shorter than $1/\Delta$.
The time resolution can be understood as the uncertainty in the time
at which the photon caused a change in the detector, due to technical
imperfections in the detector, and finite bandwidth of the
associated electronics. Note that we cannot assume truly
infinitesimal time resolution, since the quantum jumps formalism
that we apply here uses Born, Markov, and rotating wave
approximations which break down on very short timescales on the
order of the inverse of the optical frequency. Thus the results in
this section are valid in the regime $1/ \omega_{1,2} \ll t_r \ll 1
/ \Delta$.

As noted at the end of Section \ref{sec:barrett+kok}, we can analyze
the scheme as if the collection and detection efficiency was
perfect, since we are only interested in the final state in the case
when detector clicks where actually observed on both rounds of the
protocol. Finite collection and detection efficiencies will not
affect this conditional state, but will simply reduce the overall
success probability. In this case, the evolution of the system when
no detector clicks are observed is described by the Shr\"odinger
equation,
\begin{eqnarray}
\label{eqn:EffectiveSchrodingerEquation}
    \frac{\rm d}{{\rm d}t} |\psi(t) \> &=& -\frac{i}{\hbar} H_{\rm cond}
    |\psi(t)\> \, .
\end{eqnarray}
Since $H_{\rm cond}$ is non-Hermitian, this evolution is non-unitary,
and thus $|\psi(t) \>$ is unnormalized. The norm-squared of the
wavefunction, $||\psi(t) \>|^2$, can be interpreted as the
probability that the system has not emitted any photons since the
previous emission event.

In the event of a detector click in the $D_+$ or $D_-$ detectors,
the state evolves discontinuously according to
\begin{eqnarray}
\label{eqn:idealop}
    | \psi_+' \> &=& \frac{R_+ | \psi \>}{\< \psi | R_+^{\dag} R_+ | \psi \>}
    \, , \nonumber\\
    | \psi_-' \> &=& \frac{R_- | \psi \>}{\< \psi | R_-^{\dag} R_- | \psi \>}
    \, ,
\end{eqnarray}
respectively. The corresponding probability density for a click in
either detector is given by
\begin{eqnarray}
\label{eqn:probclick}
    u(t,\pm)&=& \<\psi(t)|R_{\pm}^{\dag} R_{\pm}|\psi(t)\> \, .
\end{eqnarray}
Here, $u(t,\pm)dt$ is the total probability that a click occurs in
detector $D_{\pm}$ between times $t$ and $t+dt$, and that no click
occurred before time $t$.

Following the first optical $\pi-$pulse, the state of the two-atom
system at $t=0$ is given by $\vert \psi(0) \rangle =
\frac{1}{2}(|00\> + |02\> + |20\> + |22\>)$. Note that we are
interested only in the parts of the state that will ultimately be
post-selected conditional on observing detector clicks on both
rounds of the protocol. Thus we can restrict attention to the
components of the state in the subspace spanned by the
states $|01\>$, $|10\>$, $|02\>$, and $|20\>$. This leads to the
following closed set of coupled equations for the evolution
generated by Eq. (\ref{eqn:EffectiveSchrodingerEquation}),
\begin{eqnarray}
\label{eqn:rates_ideal}
    \dot{c}_{02} &=& -i \Delta c_{02} - \frac{\kappa_{\rm eff}}{2} c_{02}\,, \nonumber\\
    \dot{c}_{20} &=& - \frac{\kappa_{\rm eff}}{2} c_{20}\,, \nonumber\\
    \dot{c}_{01} &=& 0 \,, \nonumber\\
    \dot{c}_{10} &=& 0 \,, \nonumber\\
\end{eqnarray}
where $c_{jk} (t)  = \langle j \, k \vert \psi(t)\rangle$.
Eqs.~(\ref{eqn:rates_ideal}) may be readily solved to give
\begin{eqnarray}
\label{eqn:ratesSoln_ideal}
    c_{02}(t) &=& c_{02}(0) \e^{-(i \Delta + \kappa_{\rm eff}/2)t} \, ,\nonumber\\
    c_{20}(t) &=& c_{20}(0) \e^{-\kappa_{\rm eff} t/2 } \, . \nonumber\\
\end{eqnarray}
Since $c_{02;00}(0) = c_{20;00}(0)$, we see that these two
coefficients are identical up to a varying phase factor,
\begin{eqnarray}
\label{eqn:ratesPhase_ideal}
    c_{02} &=& \e^{-i \Delta t} c_{20} \, .
\end{eqnarray}
By applying Eq.~(\ref{eqn:idealop}), we find that the normalized
zero-excitation component of the state after the first click  at
$t_1$ is given by
\begin{eqnarray}
\label{eqn:stateClick1_ideal}
    | \psi_{\pm}(t_1) \> &=& \frac{|01\> \pm \e^{i \Delta t_1}|10\>}{\sqrt{2}}
    \, ,
\end{eqnarray}
corresponding to a click in the $D_+$ or $D_-$ detector
respectively. Note that the true state will include terms
proportional to $|12\>$ and $|21\>$. However, these will be
post-selected away on the second round of the protocol, and so can
be neglected for the purposes of this analysis. Proceeding with the
second round as described in Section \ref{sec:barrett+kok}, we
find that the second click at $t_2$ removes these unwanted terms,
and leaves the system in
\begin{eqnarray}
\label{eqn:stateClick2_ideal}
    | \psi(t_1,t_2) \> &=& \frac{|01\> + (-1)^{m}
    \e^{i \Delta (t_1-t_2)}|10\>}{\sqrt{2}} \, ,
\end{eqnarray}
where $m$ is the number of $D_-$ clicks that have been observed. We
note that regardless of the actual values of $t_1$ and $t_2$, the
final state for any detector click combination is always a maximally
entangled state. In addition, as long as the values $t_1$ and $t_2$
are known, it is possible in principle to undo this additional phase
shift in the system with a local operation. This may be achieved via
a rotation about the $z$-axis by an angle $-\Delta (t_1-t_2)$ on the
first qubit. Therefore in this ideal case the fidelity of the
protocol is unity despite mismatch of the cavity frequencies.

This result can be understood as follows. In general we might expect
a reduction in fidelity due to the fact that the frequency of the
photons carries some `which path' information about which of the
atoms the detected photons originated from. However, since we have
assumed idealized time-resolution detectors, complementary
information about the frequency of the photon cannot be determined,
even in principle. Thus these idealized detectors themselves erase
the `which path' information, contained in the frequencies of the photons,
which might otherwise have reduced the fidelity.

\subsection{Bad detector limit}
\label{sec:bad_ent}

We now consider the case of very bad time resolution detectors, when
$t_r \gg \Delta^{-1},~ \kappa_{\rm eff}^{-1}$. In this regime, the
detectors give essentially no information about the arrival time of
the photons, but simply indicate whether a photon was observed or
not on a given round of the protocol. We again assume that the
collection and detection efficiencies are unity.

By inspecting Eq.~(\ref{eqn:stateClick2_ideal}) it is clear that,
since $t_1$ and $t_2$ will now be unknown, an unknown phase will be
accumulated between the two terms in the superposition. Averaging
over this phase will lead to a mixed state, with less than ideal
fidelity. Assuming the click is observed in detector $D_+$, the
state at the end of the first round of the protocol will be
\begin{eqnarray}
\label{eqn:badlimit_1clickrho}
    \bar{\rho}_+ &=&
    \frac{1}{p(+)} \int_0^\infty dt_1
    p(t_1,+) \nonumber\\
    && \times \frac{1}{2}    (|10\> + \e^{-i \Delta t_1} |01\>)
    (\<10| + \e^{i \Delta t_1} \<01|) \,. ~~
\end{eqnarray}
Here, $p(t_1,+)dt$ is the probability that a single photon is
emitted into $D_+$ between times $t_1$ and $t_1+dt$, and that no
photon is observed subsequently in  either detector. $p(+) =
\int_0^\infty dt_1 p(t_1,+)$ is the total probability of observing
precisely one photon in $D_+$, and no photons in  $D_-$. These
quantities can be calculated within the QJ method to give
\begin{equation}
p(t_1,+)  = \frac{\kappa_{\rm eff} \e^{-\kappa_{\rm eff}t_1}}{4}\,,
\end{equation}
from which we obtain $p(+) = 1/4$. The state of the system at the
end of the first round is found to be
\begin{eqnarray}
\label{eqn:badlimit_1clickrho_answer}
    \bar{\rho}_+ &=&
    \frac{1}{2}\Big(
    |01\>\<01| + |10\>\<10| \nonumber\\
    && + \frac{\kappa_{\rm eff}}{\kappa_{\rm eff} + i \Delta} |01\>\<10|
    + \frac{\kappa_{\rm eff}}{\kappa_{\rm eff} - i \Delta} |10\>\<01|
    \Big )\,. ~~~
\end{eqnarray}
To determine the effect of the second round of the protocol, we use
$\bar{\rho}_+$ as the input, and evolve under the generalized
Schr\"odinger equation, $\dot{\rho}(t_2) = -{\textstyle
\frac{i}{\hbar}}[H_{\rm cond} \rho(t_2) - \rho(t_2) H_{\rm
cond}^{\dag}]$. The state corresponding to a click in $D_+$ at time
$t_2$ is then $R_{+} \bar{\rho}_{+}(t_2) R_{+}^{\dag} / \mathrm{tr}
[R_{+} \bar{\rho}_{+}(t_2) R_{+}^{\dag}]$, and the corresponding
probability density is $p(t_2,+) = \mathrm{tr} [R_{+}
\bar{\rho}_{+}(t_2) R_{+}^{\dag}]$. From this we find that the final
state, after both rounds of the protocol, is given by
\begin{eqnarray}
\label{eqn:badlimit_2clickrho}
    \bar{\rho}_{++} &=&
    \int_0^{\infty} dt_2 {R_{+} \bar{\rho}_{+}(t_2)
    R_{+}^{\dag}}
     \, ,
    \nonumber\\
    &=&
    \frac{1}{2}\Big(
    |01\>\<01| + |10\>\<10| \nonumber\\
    && + \frac{\kappa_{\rm eff}^2}{\kappa_{\rm eff}^2 + \Delta^2} (|01\>\<10|
    + |10\>\<01|)   \Big ) \,.
\end{eqnarray}

\begin{figure}
\begin{center}
\resizebox{\columnwidth}{!}{{\includegraphics {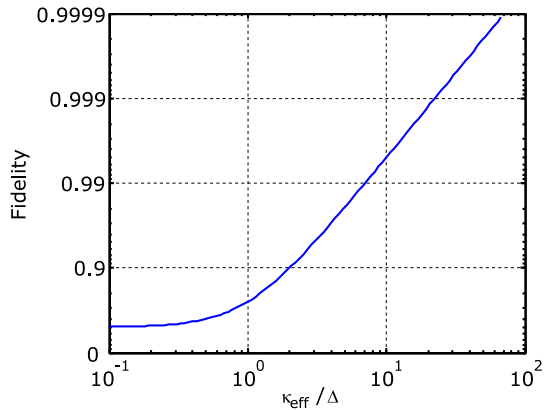}}}
\end{center}
\vspace*{-0.5cm}
\caption{Log-log plot of the final state fidelity as a function of $\kappa_{\rm eff} / \Delta$. We notice that even for a bad detector which is in principle able to distinguish the different frequency photons, a very high fidelity is possible provided that the bandwidth of the photons is large compared with their detuning.} \label{fig:fidbaddetector}
\end{figure}

A useful figure of merit for evaluating the effect of frequency
mismatch is the fidelity of the final state with the closest
maximally entangled state. Note that the protocol is such that the
final state only has components in the $\{|01\>,|10\>\}$ basis, so
the closest maximally entangled state will be of the form
$|\psi(\phi)\> = (|01\> + \e^{i \phi} |10\>)/\sqrt{2}$. Thus, for a
given state $\rho$, the fidelity can be found by maximizing
\begin{eqnarray}
\label{eqn:FidDef}
    F(\rho,\phi) = \< \psi(\phi) | \rho | \psi(\phi) \> \, ,
\end{eqnarray}
with respect to $\phi$, i.e. solving $\partial F(\rho,\phi) /
\partial \phi = 0$. Doing so we find that $\phi = -\arg(\rho_{01,10})$, and that the
fidelity is given by
\begin{eqnarray}
\label{eqn:Fidt1t2}
    F(\rho) &=& \frac{1}{2} + |\rho_{01,10}| \,.
\end{eqnarray}
Substituting $\bar{\rho}_{++}$ from Eq.
(\ref{eqn:badlimit_2clickrho}) gives
\begin{eqnarray}
\label{eqn:Fidt1t2ExplicitExpression}
    F(\bar{\rho}_{++}) &=& \frac{1}{2}\left( 1 + \frac{\kappa_{\rm eff}^2}{\kappa_{\rm eff}^2 + \Delta^2}\right) \,.
\end{eqnarray}
We plot this result in Figure \ref{fig:fidbaddetector}.
$F(\bar{\rho}_{++})$ increases toward unity for $\kappa_{\rm eff}
\gg \Delta$. This result can be understood by considering the
spectrum of the emitted photons. More specifically, $\kappa_{\rm eff}$ can be thought
of as the bandwidth of each photon in the frequency domain. Thus
for $\kappa_{\rm eff} \gg \Delta$, the photons become spectrally
indistinguishable, and therefore carry no which-path information. In this
regime we can obtain a large fidelity, even with very low
time-resolution detectors.

\subsection{Intermediate detector case}
\label{sec:moderate_ent}

In this section we describe the effect of having a detector with
arbitrary time resolution, which may lie between the two limiting
cases already studied in the preceding sections. This requires a
more sophisticated model of the detection process to include a
finite time resolution.

\begin{figure}
\begin{center}
\resizebox{\columnwidth}{!}{{\includegraphics {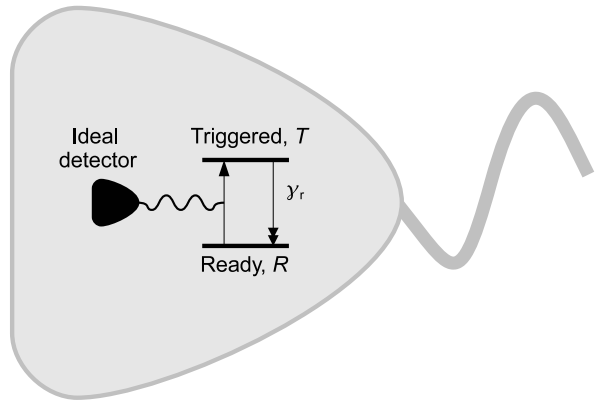}}}
\end{center}
\vspace*{-0.5cm}
\caption{Diagram showing the detector model adopted here to take account of the finite time resolution of the detector.} \label{fig:detector}
\end{figure}

The detector model we use is similar to the one described in
Refs.~\cite{wiseman,wiseman2}, and is summarized in Figure
\ref{fig:detector}. While the complete model in
Refs.~\cite{wiseman,wiseman2} incorporates a number of realistic
imperfections, including finite time resolution, dead time, and dark
counts, here we restrict our attention to the effect of finite time
resolution alone.

We now briefly review the model of \cite{wiseman,wiseman2} for the
case of a single fluorescing quantum system, observed by a single
detector, neglecting dead time and dark counts. In this case
\cite{wiseman,wiseman2}, the model consists of an idealized detector
coupled to the transition between two internal states, {\em ready}
and {\em triggered} (denoted here by $R$ and $T$, respectively), of
the detector. When a photon is registered by the ideal detector, the
internal state changes from $R$ to $T$. The $T$ state of the
detector decays, via a Poisson process, back to $R$ at a rate
$\gamma_r$. This decay process can be understood as the observed
detector `click', i.e. the time of the $T \to R$ transition
corresponds to the photon detection time reported to the observer.
$\gamma_r^{-1}$ can be understood as the response time of the
detector.

The quantities of interest are the unnormalized density matrices
$\rho_R(t)$ and $\rho_T(t)$. $\rho_R(t)$ corresponds to the state of
the quantum system at time $t$, given that the detector is in the
`ready' state, while $\rho_T(t)$ corresponds to the state of the
quantum system given that the detector is in the `triggered' state.
The states are unnormalized, such that
$\mathrm{tr}[\rho_R(t)+\rho_T(t)]$ is the total probability that no
click has been observed up to time $t$ (provided the state was
normalized at time $t=0$). The normalized state of the system at
arbitrary times is then given by $\rho(t) = [\rho_R(t)+\rho_T(t)] /
\mathrm{tr}[\rho_R(t)+\rho_T(t)]$.

Between detector clicks, the state evolves according to the
generalized mater equation
\begin{eqnarray}
\label{eqn:detectorevolutionSingleDetector}
    \dot{\rho}_{R} &=& ({\cal L}  - J[R])\rho_{R} \, , \nonumber\\
    \dot{\rho}_{T} &=& ({\cal L} - \gamma_r)\rho_{T} + J[R] \rho_{R} \, ,
\end{eqnarray}
where ${\cal L} \rho \equiv {\textstyle -\frac{i}{\hbar}}(H_{\rm
cond} \rho - \rho H_{\rm cond}^{\dag}) + R \rho R^{\dag}$ is the
usual Lindblad superoperator, and $J[R]\rho \equiv R \rho R^{\dag}$
is the jump superoperator. $R$ here is the appropriate lowering
operator for the fluorescing system of interest, with $H_{\rm cond}$
the corresponding conditional, non-Hermitian Hamiltonian.

The first equation corresponds to the conditional evolution for the
case when no photon is emitted, and is a generalization of Eq.
(\ref{eqn:EffectiveSchrodingerEquation}) to mixed states. The second
equation has three contributions. The term ${\cal L}\rho_{T}$
corresponds to the unconditional evolution of the system, i.e. it is
what would be expected if there were no detector present. This can
be understood since, once the detector is in the triggered state, no
more information about the state of the system can be obtained from
the detector until it returns to the `ready' state via a `click'
event. The term proportional to $\gamma_r \rho_{T}$ ensures that
$\rho_{T}$ is appropriately normalized, given that no click has been
observed. Finally, the $J[R] \rho_{R}$ term couples the two
equations and corresponds to  $R$ to $T$ transitions of the detector
when a photon is received.

The probability distribution for observing a detector click at time
$t$, given that no clicks were observed until time $t$, and provided
the state was normalized at time $t=0$, is given by
\begin{equation}
p(t) = \gamma_r \mathrm{tr}[\rho_T(t)] \,.
\end{equation}
When a click occurs, the state evolves discontinuously as
\begin{align}
\label{eqn:detectorevolutionSingleDetectorJump}
\rho_R(t + dt) & =  \rho_T(t) \, ,\nonumber\\
\rho_T(t + dt) & = 0 \, .
\end{align}
Eqs.~(\ref{eqn:detectorevolutionSingleDetector}--\ref{eqn:detectorevolutionSingleDetectorJump})
allow one to calculate the probability distributions for click times,
and also to determine the state of the system at arbitrary times,
conditional on the record of detector click times.

It is straightforward to generalize this model to the case of two
detectors, observing two atoms via a beam splitter. We now have to
track four quantities, $\rho_{RR}$, $\rho_{+}$, $\rho_{-}$, and
$\rho_{TT}$. $\rho_{RR}$ corresponds to the state of the two atom
system when both detectors are in the ready state. $\rho_+$
($\rho_-$) corresponds to the detector $D_+$ in the triggered
(ready) state, and $D_-$ in the ready (triggered) state. $\rho_{TT}$
corresponds to both detectors being in the triggered state.

The set of equations governing the evolution of the system
conditioned on {\em no detector click} is
\begin{eqnarray}
\label{eqn:detectorevolution}
    \dot{\rho}_{RR} &=& ({\cal L}  - J[R_{+}] - J[R_{-}])\rho_{RR} \, ,\nonumber\\
    \dot{\rho}_+ &=& ({\cal L}  - J[R_{-}] - \gamma_r)\rho_+ + J[R_{+}]\rho_{RR}
    \, , \nonumber\\
    \dot{\rho}_- &=& ({\cal L}  - J[R_{+}] - \gamma_r)\rho_- + J[R_{-}]\rho_{RR}
    \, ,  \nonumber\\
    \dot{\rho}_{TT} &=& ({\cal L} - 2 \gamma_r)\rho_{+-} + J[R_{-}]\rho_{+} + J[R_{+}]\rho_{-} \, ,~~~
\end{eqnarray}
where ${\cal L} \rho \equiv {\textstyle -\frac{i}{\hbar}}(H_{\rm
cond} \rho - \rho H_{\rm cond}^{\dag}) + R_+ \rho R_+^{\dag}+ R_- \rho
R_-^{\dag}$ is the Lindblad master equation, $J[R]\rho \equiv R \rho
R^{\dag}$ is the jump matrix, and $\gamma_r$ is the stochastic decay
rate at which a triggered detector will signal a click and return
into the ready state.

The complimentary (instantaneous) evolution corresponding to a $D_+$
click event is now
\begin{align}
\label{eqn:detectorevolutionJumpPlus}
\rho_{RR}(t + dt) & = \rho_+(t) \, ,\nonumber\\
\rho_+(t + dt)    & = 0 \, ,\nonumber \\
\rho_-(t + dt)    & = \rho_{TT}(t) \, ,\nonumber\\
\rho_{TT}(t)      & = 0 \,.
\end{align}
while a $D_-$ click event is described by
\begin{align}
\label{eqn:detectorevolutionJumpMinus}
\rho_{RR}(t + dt) & = \rho_-(t) \, ,\nonumber\\
\rho_+(t + dt)    & = \rho_{TT}(t) \, ,\nonumber \\
\rho_-(t + dt)    & = 0 \, ,\nonumber\\
\rho_{TT}(t)      & = 0 \, .
\end{align}
We may solve the coupled equations of (\ref{eqn:detectorevolution})
analytically in the over damped regime, where $H_{\rm cond}$ and
$R_\pm$ are given by Eqs. (\ref{eqn:Heff} - \ref{eqn:jumpeff}). The
analysis can be considerably simplified by noting that we are only
interested in the case where a single detector click is observed on
each round of the protocol, since other outcomes will be discarded
as failures anyway. This means that we need only to explicitly keep
track of terms in the state which contain a single excitation.

Evaluating the Eqs.~(\ref{eqn:detectorevolution}--\ref{eqn:detectorevolutionJumpMinus}),
for the case when exactly one detector click is observed on each
round of the protocol, leads to the final state of the system given
by
\begin{widetext}
\begin{eqnarray}
\label{eqn:soln_rhos}
\rho_{01,01}(t_1,t_2) &=&
    \frac{\rho_{20,20;0}(0) \kappa_{\rm eff}^2}{4(\kappa_{\rm eff} - \gamma_r)^2}
    \e^{-\gamma_r(t_1 + t_2)}
    (\e^{- (\kappa_{\rm eff}-\gamma_r) t_1 }-1)
    (\e^{- (\kappa_{\rm eff}-\gamma_r) t_2 }-1) \, , \nonumber\\
\rho_{01,10}(t_1,t_2) &=&
    \frac{(-1)^m \rho_{20,02;0}(0) \kappa_{\rm eff}^2}
    {4(i \Delta + \kappa_{\rm eff} - \gamma_r)^2}
    \e^{-\gamma_r(t_1 + t_2)}
    (\e^{- (i \Delta + \kappa_{\rm eff} - \gamma_r ) t_1}-1)
    (\e^{- (i \Delta + \kappa_{\rm eff}  - \gamma_r) t_2}-1) \, ,\nonumber\\
\rho_{10,10}(t_1,t_2) &=&
    \frac{\rho_{02,02;0}(0) \kappa_{\rm eff}^2}{4(\kappa_{\rm eff} - \gamma_r)^2}
    \e^{-\gamma_r(t_1 + t_2)}
    (\e^{- (\kappa_{\rm eff}-\gamma_r) t_1 }-1)
    (\e^{- (\kappa_{\rm eff}-\gamma_r) t_2 }-1) \, ,
\end{eqnarray}
\end{widetext}
where $t_1$ and $t_2$ are the times of the observed detector clicks
on the corresponding rounds of the protocol, and $m=0$ if both
clicks are in the same detector, else $m=1$. All other
components of the density matrix vanish. Note that, as before, this state is
unnormalized, with the normalization such that the joint probability
distribution for any specific two click combination is given by $P(t_1,t_2) \equiv \gamma_r^2 \mathrm{tr}[\rho(t_1,t_2)]$. As expected, this expression integrates to ${\textstyle \frac{1}{8}}$ which is the total probability of having exactly two clicks in a specific sequence such as $D_+$ followed by $D_+$.

For $\Delta \gg \kappa_{\rm eff},~\gamma_r$ we see that the
coherence vanishes corresponding to a large amount of mixing, which
confirms the results from Section \ref{sec:bad_ent}. Similarly, for
$\Delta \ll \kappa_{\rm eff},~\gamma_r$, we find that we recover the
pure, maximally entangled state from Section \ref{sec:ideal_ent}.

\begin{figure}
\begin{center}
\resizebox{\columnwidth}{!}{{\includegraphics {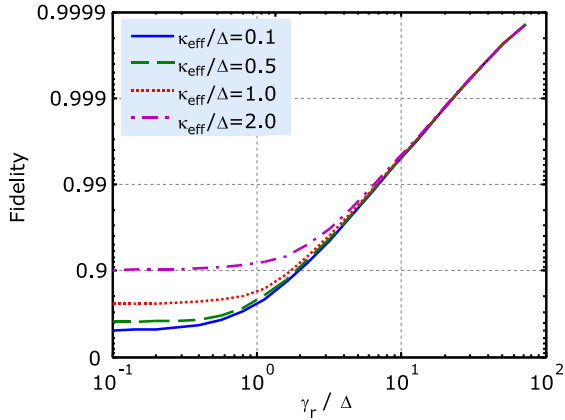}}}
\end{center}
\vspace*{-0.5cm} \caption{(Colour online) Average final fidelity for
the intermediate detector regime, from a numerical evaluation of Eq.~(\ref{eqn:FidAv2}).} \label{fig:FidAvloglog}
\end{figure}

\begin{figure}
\begin{center}
\resizebox{\columnwidth}{!}{{\includegraphics {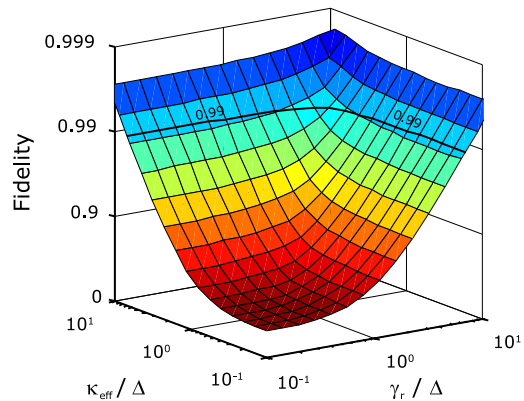}}}
\end{center}
\vspace*{-0.5cm} \caption{(Colour online) Average final fidelity for
the intermediate detector regime, from a numerical evaluation of Eq.~(\ref{eqn:FidAv2}), as a function of both $\kappa_{\rm eff}$ and
$\gamma_r$. } \label{fig:FidAv3d}
\end{figure}

More specifically, using Eq.~(\ref{eqn:Fidt1t2}) we find that the
fidelity of the protocol for this intermediate detector case is
given by
\begin{widetext}
\begin{eqnarray}
\label{eqn:Fidt1t2_intermediate}
    F(t_1, t_2) &=& \frac{1}{2} +
    0.5 \frac { ( \kappa- \gamma_r)^{2}}
    {(\kappa- \gamma_r )^2 + \Delta^2}
    \frac{|(\e^{- (  i \Delta+\kappa -\gamma_r ) t_1}- 1 )
    (\e^{- (  i \Delta+\kappa -\gamma_r) t_2}- 1 )| }
    { ( \e^{- (\kappa-\gamma_r) t_1}-1 )( \e^{- (\kappa-\gamma_r) t_2}-1 ) }
    \,.
\end{eqnarray}
\end{widetext}
The average fidelity of the protocol is then
\begin{eqnarray}
\label{eqn:FidAv}
    \bar{F} &=& \frac{\int_0^\infty {\rm d} t_1 \int_0^\infty {\rm d} t_2 P(t_1, t_2) \cdot F(t_1, t_2)}{\int_0^\infty {\rm d} t_1 \int_0^\infty {\rm d} t_2 P(t_1, t_2)}~~~.
\end{eqnarray}
Using Eqs.~(\ref{eqn:soln_rhos}) and (\ref{eqn:Fidt1t2}) and the
joint probability density for the clicks, $P(t_1,t_2)$, this simplifies to
\begin{eqnarray}
\label{eqn:FidAv2}
    \bar{F} &=& {\textstyle \frac{1}{2}} + \int_0^\infty {\rm d} t_1 \int_0^\infty {\rm d} t_2 8 \gamma_r^2 |\rho_{01,10}(t_1,t_2)| \,.
\end{eqnarray}
This expression is not readily integrable analytically, but it may
be solved numerically. Doing so for a range of $\gamma_r$
and $\kappa_{\rm eff}$ gives the results plotted in Figures
\ref{fig:FidAvloglog} and \ref{fig:FidAv3d}. Figure
\ref{fig:FidAvloglog} indicates that the error in the entangling
operation [that is, $1-\bar{F}$, with $\bar{F}$ the average fidelity
of Eq.~(\ref{eqn:FidAv2})] scales approximately quadratically with
$\Delta/\gamma_r$, and also indicates that for fidelities exceeding
0.99 we require $\gamma_r \gtrsim 6 \Delta$.  We also find that
$\kappa_{\rm eff}$ and $\gamma_r$ have similar influences on the
final state fidelity, as can be seen from the plot shown in Figure
\ref{fig:FidAv3d}.


\section{Hong-Ou-Mandel dip with detuned sources}
\label{sec:HOM}

In this section, we consider another few-photon interference
phenomenon, the Hong-Ou-Mandel (HOM) effect, with detuned photons.
The HOM effect is well understood and was first observed in 1987
\cite{HOM}. In the usual setup, two photons with identical spectra
and spatio-temporal mode shapes simultaneously impinge on the two
different input ports of a beam splitter. Provided these photons are
identical and phase coherent, the state at the output ports of the
beam splitter is then given by $\vert \psi_{\mathrm{out}} \rangle =
(\vert 2, 0 \rangle + \vert 0, 2 \rangle)/\sqrt{2}$. In this
idealized situation, subsequent detection of the photons by
photodetectors placed at each output port thus reveals perfect
\emph{bunching} or \emph{coalescence} of the photons; that is, both
detection events occur in the same detector, and coincidences
corresponding to one photon leaving each output port of the
beam splitter are not detected.

The HOM effect was originally used to characterize the coherence of
individual photons from parametric down conversion sources
\cite{HOM}, and has also been used to characterize photons from a
single quantum dot source \cite{Santori2002}. More recently, this
effect has been used as the basis of many schemes for linear
optical quantum computing \cite{KLM,KokReview}, and has been
proposed as a method for entangling spatially separated atomic
ensembles \cite{Chen06i}. The effect has recently been observed
using photons from independent sources, such as independently
trapped neutral atoms \cite{Beugnon2006} and separate ions
\cite{Maunz2006}.

The ubiquity and utility of this phenomenon in quantum information
processing therefore motivates the quantitative study of the HOM
effect when the photons are not identical, in particular when the
center frequency of the two photons is not identical. HOM
interference of such detuned photons has already been studied both
theoretically \cite{Legero2003} and in an elegant experiment using
photons from an atom-cavity system \cite{Legero2004}. Legero
\emph{et al.} predicted, and subsequently observed, that perfect
photon coalescence is not seen. Rather, provided sufficient time
resolution is available in the detector signals, the probability of
detecting both photons in the same detector oscillates as a function
of the detection time, with the frequency of this beat oscillation
given by the relative detuning of the two input photons.

Here, we study this effect using a similar method to that described
in Section \ref{sec:entangle_mismatch}, i.e.~we treat the effect as a
continuous measurement problem, and keep track of the internal
quantum state of the sources. In Section \ref{sec:HOMidealDetector} we
consider the limit of ideal detectors, namely ones for which precise
information about the timing of the detection events is available.
As noted above, a similar calculation has already been performed in
\cite{Legero2003} using different techniques. We obtain results in
agreement with Ref.~\cite{Legero2003}, which demonstrates the
validity of our approach, and furthermore, we hope sheds an
alternative perspective on the result. Then in Section
\ref{sec:HOMBadDetector} we discuss the opposite limit, and
determine the coalescence probability in the case where no timing
information is available from the detectors. We quantitatively consider the
intermediate case in Section \ref{sec:HOMIntermediateDetector}, where the detectors have finite time resolution,
and evaluate the visibility of the HOM beat fringes as a function of
the both the photon detuning and detector resolution.

\subsection{Ideal detector case}
\label{sec:HOMidealDetector}

\begin{figure}
\begin{minipage}{\columnwidth}
\begin{center}
\resizebox{\columnwidth}{!}{\rotatebox{0.0}{\includegraphics{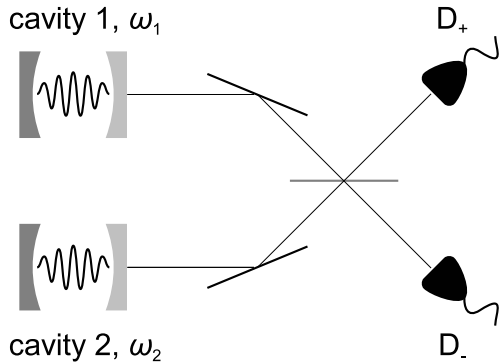}}}
\end{center}
\vspace*{-0.5cm} \caption{(Colour online) Set-up for a HOM
experiment with detuned photons. The single photon sources are each
modeled as single mode cavities, with a leaky mirror of decay rate
$\kappa$ at one end. The frequencies of the cavities (and hence the
emitted photons) are $\omega_1$ and $\omega_2 = \omega_1 - \Delta$
respectively. The sources are initially prepared in the single
photon Fock state $|1,1\rangle = c_1^\dag c_2^\dag |0,0\rangle$.
Photons emitted from these sources impinge on the two different
input ports of a beam splitter, and are subsequently detected by the
photodetectors $D_+$ and $D_-$. The cavity sources may be replaced
with a variety of different physical systems, with qualitatively
similar results, as described in the text.} \label{fig:HOMsetup}
\end{minipage}
\end{figure}

In this section we consider the HOM effect with detuned photons, of
center frequencies $\omega_1$ and $\omega_2 = \omega_1 - \Delta$ in
the limit of detectors with time resolution much shorter than
$\Delta^{-1}$. We use a simple model of the single photon sources,
comprising single mode leaky cavities, each initially prepared in
the single photon Fock state, as shown in Figure \ref{fig:HOMsetup}.

As elsewhere in this paper, we treat this setup as a continuous
measurement experiment, and apply the QJ formalism. We assume, for
simplicity, that the efficiencies for photodetection, collection,
and photon emission into the desired mode are all unity. This
assumption can be relaxed later. Provided the emission efficiencies
are identical, the results will be essentially the same, except for
an overall reduction in coincidence count rate.

The quantities of interest are the time-resolved coincidence
probability density functions, $p(t_1,t_2,\pm,\pm)$. For instance,
$p(t_1,t_2,+,-)dt_1 dt_2$ is the probability that the first click is
observed in the detector $D_+$ in the infinitesimal interval
$[t_1,t_1+dt_1]$, and that the second click is observed in $D_-$ in
the interval $[t_2,t_2+dt_2]$. We restrict our attention here to the
quantity $p(t_1,t_2, +, +)$, but the other quantities are
straightforward to calculate in a similar manner (in particular the
symmetry of the setup implies $p(t_1,t_2, +, +) = p(t_1,t_2, -,
-)$).

Between photodetection events, i.e.~conditional on no detector
clicks, the cavity systems evolve under the Schr\"{o}dinger equation
[Eq.~(\ref{eqn:EffectiveSchrodingerEquation})] according to the
non-Hermitian conditional Hamiltonian

\begin{eqnarray}
\label{eqn:HOMHeff}
    H_{\mathrm{cond}} &=& \hbar \omega_1 b_1^\dag b_1
    + \hbar \omega_2 b_2^\dag b_2
    - \frac{i \hbar \kappa}{2} b_1^\dag b_1
    - \frac{i \hbar \kappa}{2} b_2^\dag b_2 \,. ~~~~~
\end{eqnarray}
Here, $\omega_j$ is the frequency of cavity $j$ (and hence the
central frequency of the emitted photon), $b_j$ are the
corresponding lowering operators, and $\kappa$ is the leakage rate
of each cavity. For simplicity, we assume that the cavity leakage
rate, and hence the temporal mode shape of the emitted photons, is
the same for each cavity.

As in Section \ref{sec:entangle_mismatch}, in the event of a
detection event the system evolves discontinuously according to
\begin{eqnarray}
    | \psi_+' \> &=& \frac{R_+ | \psi \>}{\< \psi | R_+^{\dag} R_+ | \psi \>} \, , \label{eqn:HOMidealopPlus} \\
    | \psi_-' \> &=& \frac{R_- | \psi \>}{\< \psi | R_-^{\dag} R_- | \psi
    \>} \, . \label{eqn:HOMidealopMinus}
\end{eqnarray}
In this case, the jump operators are given directly by the beam
splitter transformation,
\begin{eqnarray}
\label{eqn:HOMjump}
    R_+ &=& \sqrt{\frac{\kappa}{2}} (b_1 + b_2) \, ,\nonumber\\
    R_- &=& \sqrt{\frac{\kappa}{2}} (b_1 - b_2) \, .
\end{eqnarray}

It is worth noting that while Eqs.~(\ref{eqn:HOMHeff}) and
(\ref{eqn:HOMjump}), together with the initial condition that each
cavity is prepared in a Fock state, represent a rather idealized
model of a single photon source, they are also directly relevant for
more realistic systems that could be used as sources. These include
isolated two-level atomic systems (such as trapped atoms, ions,
quantum dots, impurity centers in diamond), or corresponding systems
coupled to a single (adiabatically eliminated) cavity mode in the
bad-cavity regime. In such cases, the cavity mode is replaced with
the corresponding atomic transition, and the lowering operators
$b_j$ should be replaced with the corresponding lowering operator
for the atomic system, similarly to Section \ref{sec:entangle_mismatch}.
The results below therefore also apply to
these alternative systems.

We assume the cavities are initially prepared in the Fock state
$\vert \psi(0) \rangle = |1,1\rangle = b_1^\dag
b_2^\dag|0,0\rangle$, where $|0,0\rangle$ represents the vacuum mode
for both cavities. Subsequently, conditioned on no detector clicks
being observed between times $0$ and $t_1$, the state for the
cavities evolves according to $|\tilde{\psi}(t_1)\rangle =
\exp(-{\textstyle \frac{i}{\hbar}}H_{\mathrm{cond}} t_1) \vert \psi(0) \rangle =
\exp[-(i\omega_1 +i \omega_2 +\kappa )t_1]|1,1\rangle $.

Assuming a click occurs at time $t_1$ in detector $D_+$, the state
evolves according to Eq.(\ref{eqn:HOMidealopPlus}), and we find that
the normalized state after the detector click is
\begin{eqnarray}
\label{eqn:HOMStateAfterFirstClick}
    \vert \psi(t_1+dt) \rangle & = & \frac{1}{\sqrt{2}}
    \left(b_1^\dag + b_2^\dag \right) |0, 0 \rangle  \, , \nonumber\\
    & = & \frac{1}{\sqrt{2}}
    \left(|1, 0 \rangle + |0, 1 \rangle \right)\,.
\end{eqnarray}
The probability density for this detector click to occur between
times $t_1$ and $t_1+dt$, conditional on no clicks being observed up
to time $t_1$, is given by $p(t_1, +)dt = \vert\vert R_+
|\tilde{\psi}(t_1)\rangle \vert\vert^2 dt  = \kappa \e^{-2\kappa t_1}
dt$. Note that Eq.~(\ref{eqn:HOMStateAfterFirstClick}) implies that
the sources become entangled after the first detector click.
Furthermore, the state of the system immediately after this first
click is independent of both the click time $t_1$ and the detuning
of the two photons, $\Delta$.

Conditional on no clicks being observed between times $t_1+dt$ and
$t_2$, the state again evolves under the Schr\"{o}dinger equation
with $H_{\mathrm{cond}}$, as
\begin{equation}
\begin{split}
\label{eqn:HOMStateBeforeSecondClick}
    \vert \tilde\psi(t_2) \rangle = &
    \exp[{-{\textstyle \frac{i}{\hbar}} H_{\mathrm{cond}} (t_2-t_1)}]
    \vert  \psi(t_1+dt) \rangle  \, ,    \\
     = & \frac{1}{\sqrt{2}} \left(\e^{-(i \omega_1 + \frac{\kappa}{2}) (t_2-t_1)} |1,0\rangle \right.   \\
      & \qquad \qquad \qquad \left. + \, \e^{-(i \omega_2 + \frac{\kappa}{2}) (t_2-t_1)} |0,1\rangle
      \right) \, , \\
     = & \frac{\e^{-(i \omega_1 + \frac{\kappa}{2})
     (t_2-t_1)}}{\sqrt{2}} \left( |1,0\rangle  + \e^{i \Delta  (t_2-t_1)} |0,1\rangle \right) \,.
\end{split}
\end{equation}
Note that the relative phase between the two terms in this
expression oscillates in time at a rate $\Delta$. It is this
oscillation that leads to the `quantum beats' in the photon
coalescence probability.

A little algebra shows that the probability density for the second
detector click to occur in detector $D_+$, between times $t_2$ and
$t_2+dt$, conditional on the first click being observed in $D_+$ at
time $t_1$, is given by
\begin{align}
p(t_2, + | t1, + )  & = \vert\vert R_+ |\tilde{\psi}(t_2)\rangle
\vert\vert^2 \, , \nonumber\\
& = \kappa \e^{-\kappa(t_2-t_1)} \cdot \frac{1}{2} \left\{1 +
\cos[\Delta (t_2-t_1) ] \right\} \,.
\label{eqn:HOMTimeResolvedCoalescenceProbability}
\end{align}

\begin{figure}
\begin{minipage}{\columnwidth}
\begin{center}
\resizebox{\columnwidth}{!}{\rotatebox{0.0}{\includegraphics{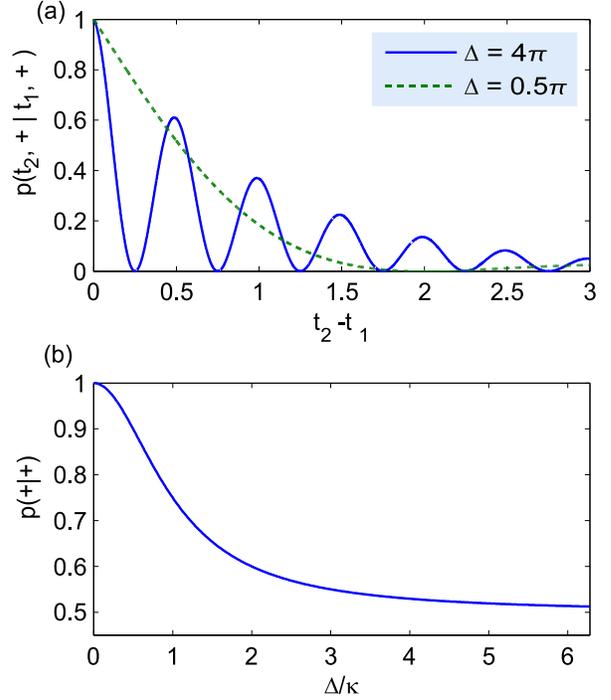}}}
\end{center}
\vspace*{-0.5cm} \caption{(Colour online) Generalized Hong-Ou-Mandel
effect with detuned photons. (a) Probability density function for
observing both clicks in the same detector, $p(t_2, + | t1, + ) =
p(t_2, - | t1, - )$, as a function of the delay between clicks,
$(t_2-t_1)$. This probability density oscillates at a rate $\Delta$,
the detuning between the center frequencies of the two photons.
$\kappa = 1$, $\Delta = 4 \pi$ (solid curve), $\Delta = 0.5 \pi$
(broken curve). (b) Total coalescence probability, $p(+|+) = p(-|-)$
as a function of detuning, $\Delta/\kappa$. This corresponds to the
coincidence signal that can be measured if no timing information is
available from the detectors. } \label{fig:HOMDetuning}
\end{minipage}
\end{figure}

Combining this with the expression for $p(t_1, +)$ given above we
find that the total joint probability distribution for two clicks in
detector $D_+$ is given by $p(t_1,t_2, +, +) = p(t_2, + | t1, + )
p(t_1, +) = \kappa^2 \e^{-\kappa(t_2+t_1)} \left\{1 + \cos[\Delta
(t_2-t_1) ] \right\}/2$. The somewhat more more instructive
expression $p(t_2, + | t1, + )$ is plotted as a function of
$t_2-t_1$ in Figure \ref{fig:HOMDetuning} (a).

The central result of this section is that the probability of
finding both photons in the same detector oscillates with frequency
$\Delta$, the detuning of the two photons. This is in agreement with
corresponding theoretical and experimental results in
Refs.~\cite{Legero2003} and \cite{Legero2004}, although we have used
a different method to arrive at the result \footnote{Note that our
expressions for the time resolved photon coalescence probability
density differ slightly from those considered in
Ref.~\cite{Legero2003}. The differences are due to the fact that we
consider exponentially decaying photon temporal wavefunctions,
arising from a sudden excitation of the source at time $t=0$
followed by spontaneous emission, whereas Ref.~\cite{Legero2003}
considers Gaussian temporal wavefunctions.}.

\subsection{Bad detector limit}
\label{sec:HOMBadDetector}

From the results of the previous section, it is straightforward to
calculate the total photon coalescence probability - that is, the
total probability that both photons will arrive in the same
detector. This can also be thought of as the visibility of the
Hong-Ou-Mandel effect in the limit of bad photodetectors, i.e.~detectors whose time resolution is much longer than the inverse of
the detuning, $\Delta^{-1}$. We assume that in other respects, the
detectors are ideal, in particular that their detection efficiency
is unity, and no photons are missed due to dead time. In this
limit, the only information available from the detectors is the
total number of photons arriving at each detector, and no
information about the time of arrival is available.

The total photon coalescence probability, $p(+|+)=p(-|-)$ for
observing both clicks in detector $D_+$ is given by integrating
Eq.~(\ref{eqn:HOMTimeResolvedCoalescenceProbability}) over all
values of $\tau = t_2-t_1$,
\begin{align}
p(+|+) & = \int_0^\infty d\tau p(t_2, + | t1, + ) \,, \nonumber \\
       & = \int_0^\infty d\tau \kappa \e^{-\kappa\tau} \cdot \frac{1}{2} \left\{1 +
\cos \Delta \tau  \right\}\,, \nonumber \\
       & = \frac{1}{2} \left(1 + \frac{\kappa^2}{\kappa^2 +
       \Delta^2}\right)\,.
\end{align}


This expression is plotted in Figure \ref{fig:HOMDetuning}(b). For
small detuning, $\Delta \ll \kappa$, $p(+|+)$ approaches unity,
indicating near perfect Hong-Ou-Mandel photon coalescence. For
$\Delta \gg \kappa$, $p(+|+) \simeq 1/2$, indicating that the
photons apparently act as independent particles, scattering randomly
into both output ports of the beam splitter. $\kappa$ can be thought
of as the bandwidth of the photons, giving the spread in frequency
of each photon about its center value. Thus these results agree with
the usual intuition that HOM interference can not be observed
between distinguishable photons. However, it is important to note that this is a
consequence of the poor time resolution of the detectors. As we saw
in the Section \ref{sec:HOMidealDetector}, with sufficiently fast
time resolved detectors, a modified form of the HOM interference can
be observed.

\subsection{Intermediate detector case}
\label{sec:HOMIntermediateDetector}

In Sections \ref{sec:HOMidealDetector} and \ref{sec:HOMBadDetector}
we discussed the HOM effect in the limiting cases of perfect time
resolution photodetectors, and the limit of bad detectors (with no
time resolution), respectively. In general, experiments may be in
neither limit, and the detectors may have a finite time resolution
which is comparable to the inverse of the detuning between the
incident photons. In view of the applications of the HOM effect in
quantum information processing, it is useful to have a quantitative
understanding of the effect of finite time resolution detectors. In
this section, we apply the model of finite time resolution detection
introduced in \ref{sec:moderate_ent} to analyse the HOM effect in
this regime.

As in Section \ref{sec:moderate_ent} we model the detector by the set of
equations
\begin{eqnarray}
\label{eqn:HOMdetectorevolution}
    \dot{\rho}_{RR} &=& ({\cal L}  - J[R_{+}] - J[R_{-}])\rho_{RR} \, , \nonumber\\
    \dot{\rho}_+ &=& ({\cal L}  - J[R_{-}] - \gamma_r)\rho_+ + J[R_{+}]\rho_{RR}
    \, , \nonumber\\
    \dot{\rho}_- &=& ({\cal L}  - J[R_{+}] - \gamma_r)\rho_- + J[R_{-}]\rho_{RR}
    \, , \nonumber\\
    \dot{\rho}_{TT} &=& ({\cal L} - 2 \gamma_r)\rho_{TT} + J[R_{-}]\rho_{+} + J[R_{+}]\rho_{-} \, ,~~
\end{eqnarray}
where, as before, ${\cal L} \rho \equiv {\textstyle
-\frac{i}{\hbar}}(H_{\rm cond} \rho - \rho H_{\rm cond}^{\dag}) +
R_+ \rho R_+^{\dag}+ R_- \rho R_-^{\dag}$ is the Lindblad master
equation, $J[R]\rho \equiv R \rho R^{\dag}$ is the jump matrix, and
$\gamma_r$ is the stochastic decay rate at which a triggered
detector will signal a click and return into the ready state. Now,
however, the system under observation is that shown in Figure
\ref{fig:HOMsetup}, with $H_{\rm cond}$ is given by Eq.~(\ref{eqn:HOMHeff}), and the operators $R_\pm$ are given by Eq.~(\ref{eqn:HOMjump}). Thus the matrices $\rho_{RR}$, $\rho_\pm$,
$\rho_{TT}$ describe the internal states of the two single mode
cavities representing the sources. As in Section
\ref{sec:HOMidealDetector}, we take the initial state of the system
to be the Fock state $|1,1\rangle$, such that $\rho_{RR}(0) = |1,
1\rangle \langle 1,1 \vert$, and $\rho_\pm(0) = \rho_{TT}(0) = 0$,
indicating that both detectors are in the `Ready' (R) state at time
$t=0$.

We are interested in quantities such as $p_\gamma(t_1,t_2, +, +)$,
which is the probability density function for pairs of
\emph{observed} detector clicks. This can be found by first solving Eqs.~(\ref{eqn:HOMdetectorevolution}) to find $\rho_j(t_1)$. At
a given time $t_1$, the total probability for detector $D_+$ to be
in the `Triggered' (T) state is ${\rm Tr} [\rho_+(t_1) +
\rho_{TT}(t_1)]$. Thus the probability density for the first $D_+$
detector click is given by
\begin{equation}
p_\gamma(t_1,+) = \gamma_r \, {\rm Tr} [\rho_+(t_1) +
\rho_{TT}(t_1)]\,. \label{eqn:HOMpdfFirstClick}
\end{equation}

After the first $D_+$ detector click, the state of the system is
updated and renormalized according to
\begin{align}
\rho_{RR}(t_1+dt) & =  \frac{\rho_+(t_1)}{{\rm Tr} [\rho_+(t_1) + \rho_{TT}(t_1)]}
\, , \nonumber\\
\rho_+(t_1+dt) & = 0 \, , \nonumber\\
\rho_-(t_1+dt) & = \frac{\rho_{TT}(t_1)}{{\rm Tr} [\rho_+(t_1)
    + \rho_{TT}(t_1)]} \, ,\nonumber\\
\rho_{TT}(t_1+dt) & = 0 \, .
\end{align}
This state is then used as an initial condition to Eqs.~\ref{eqn:HOMdetectorevolution}, which can be solved to give
$\rho_j(t_2|t_1,+)$, i.e.~the state of the system at time $t_2$,
conditional on a $D_+$ click at time $t_1$, and no clicks between
times $t_1$ and $t_2$. The conditional probability density for the
second click, in detector $D_+$, is then given by
\begin{equation}
p_\gamma(t_2,+|t_1,+) = \gamma_r {\rm Tr} [\rho_+(t_2|t_1,+) +
\rho_{TT}(t_2|t_1,+)] \,. \label{eqn:HOMpdfSecondClick}
\end{equation}
The joint probability distribution for $D_+$ clicks at time $t_1$
and $t_2$ can then be formed from Eqs.~(\ref{eqn:HOMpdfFirstClick})
and (\ref{eqn:HOMpdfSecondClick}) as $p_\gamma(t_1,t_2, +, +) =
p_\gamma(t_2,+|t_1,+) p_\gamma(t_1,+)$. Note that, unlike in the
ideal detector case described in Section \ref{sec:HOMidealDetector},
$p_\gamma(t_2,+|t_1,+)$ is no longer just a function of $t_2-t_1$,
but in fact depends on the value of $t_1$.

\begin{figure}
\begin{minipage}{\columnwidth}
\begin{center}
\resizebox{\columnwidth}{!}{\rotatebox{0.0}{\includegraphics{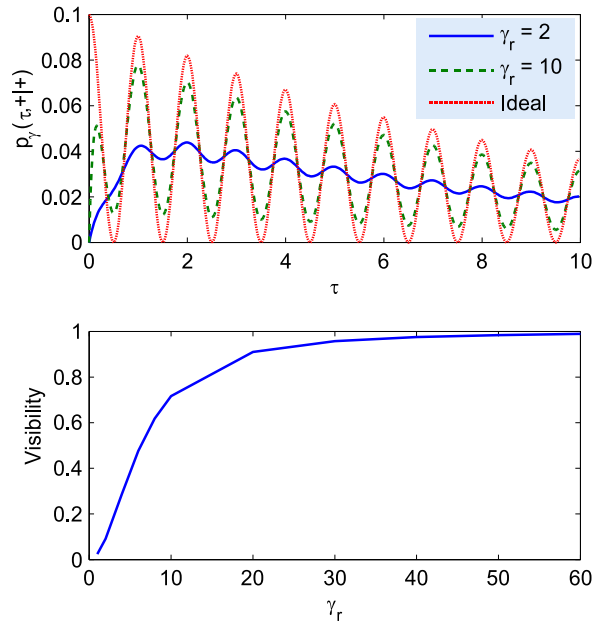}}}
\end{center}
\vspace*{-0.5cm} \caption{(Colour online) Generalized Hong-Ou-Mandel
effect with detuned photons and finite time response detectors. (a)
Probability density function for observing both clicks in the same
detector, $p_\gamma(\tau, + | +)$, as a function of the delay
between clicks, $\tau = t_2-t_1$. $\kappa = 0.1$, $\Delta = 2\pi$,
$\gamma_r=2$ (solid curve), $\gamma_r=10$ (broken curve). The dotted
curve shows the result corresponding to ideal detectors, i.e.~$p(t_2, + | t1, + )$ from Eq.~(\ref{eqn:HOMTimeResolvedCoalescenceProbability}), for the same values of $\kappa$ and $\Delta$ .
 (b) Fringe visibility at $\tau = 20 \times \pi / \Delta$ as a function of detector response rate, $\gamma_r$, for the same $\kappa$ and $\Delta$ as in (a). } \label{fig:HOMDetuningFiniteResponseDetectors}
\end{minipage}
\end{figure}

In principle, quantities such as $p_\gamma(t_1,t_2, +, +)$ can be
obtained analytically by first finding the general solution of
Eqs.~(\ref{eqn:HOMdetectorevolution}) and then applying the above
steps. However, this turns out to be difficult in general. Unlike
the case of the distant-atom entangling scheme presented in Section
\ref{sec:bad_ent}, it is not possible to concentrate on the single
photon terms in the expressions for $\rho_j$. Instead, all terms
must be tracked, including in particular those which correspond to
both photons entering the same detector within a short interval.
This means that all 64 coupled equations in
Eqs.~(\ref{eqn:HOMdetectorevolution}) must be solved. Owing to this
complexity, we resort to a numerical solution of
Eqs.~(\ref{eqn:HOMdetectorevolution}), using the {\sc Matlab}
`ode45' function. Following the steps outlined above, we obtain
$p_\gamma(t_1,t_2, +, +)$. From this quantity, we can determine
$p_\gamma(\tau, + | +)$, that is, the distribution of intervals
$\tau = t_2-t_1$ between detector clicks, given that the first click
was observed in detector $D_+$. This is the analogous quantity to
$p(t_2, + | t1, + )$, in the case of ideal detectors, which was
plotted in Figure \ref{fig:HOMDetuning} (a). $p_\gamma(\tau, + | +)$
is given by
\begin{equation}
p_\gamma(\tau, + | +) = \frac{\int_0^\infty dt_1
p_\gamma(t_1,t_1+\tau,+,+)}{\int_0^\infty dt_1 p_\gamma(t_1,+)}\,.
\end{equation}
We can also calculate $p_\gamma(\tau, - | +)$, which is the
probability distribution for the second click to occur in $D_-$,
given that the first was in $D_+$, in a similar way. In practice, we
set the upper limit of the integrals to a value of $3 \kappa^{-1}$,
which leads to a slight relative underestimate of approximately
$\exp[-2 \kappa \times 3 \kappa^{-1} ] \approx 0.25 \%$ in the value
of these integrals. $p_\gamma(\tau, + | +)$ is plotted in Figure
\ref{fig:HOMDetuningFiniteResponseDetectors}(a), along with the
ideal result, $p(t_2, + | t1, + )$ from Eq.~(\ref{eqn:HOMTimeResolvedCoalescenceProbability}), for the same
values of $\Delta$ and $\kappa$. For smaller values of the detector
response rate, $\gamma_r$, it can be seen that the interference
fringes become washed out. This reduction in visibility is a result
of the finite time response of the detectors. More specifically, as $\gamma_r$
approaches $\Delta$ it becomes harder to resolve the the individual
interference fringes.

A reasonable figure of merit for this generalized HOM effect is the
fringe visibility,
\begin{equation}
v(\tau) = \left|\frac{p_\gamma(\tau, + | +)-p_\gamma(\tau, - |
+)}{p_\gamma(\tau, + | +)+p_\gamma(\tau, - | +)}\right|\,.
\end{equation}
After an initial transient behaviour on short timescales $\tau \sim
\gamma_r^{-1}$, $v(\tau)$ is, to a good approximation, a periodic
function, with maxima at integer multiples of $\pi / \Delta$. A good
figure of merit is therefore the value of these maxima (taken
outside the initial transient region). In Figure
\ref{fig:HOMDetuningFiniteResponseDetectors}(b) we plot the fringe
visibility evaluated at $\tau = 20 \times \pi / \Delta$, as a
function of detector response rate, $\gamma_r$, for a particular
choice of $\Delta$ and $\kappa$. As expected, the fringe visibility
vanishes for slow detectors, and approaches unity for fast detectors
with $\gamma_r \gg \Delta$.

\section{Outlook and conclusions}
\label{sec:conc}

In this paper we have looked at the effect of frequency mismatch in
two schemes which make use of few-photon interference. Our main
conclusions can be summarized as follows.

We first looked at the effect of frequency mismatch in the
distributed QIP scheme of Barrett and Kok \cite{BarrettKok2005}. One
key finding was that, with idealized detectors (i.e.~such that the
detector response rate $\gamma_r$ is much faster than the photon
detuning $\Delta$) perfect entanglement of distant atoms is still
possible. The entangled states pick up a known phase that
depends on the times of the photodetection events, and which can
be corrected via a local unitary operation. 
This is true as long as the cavity mismatch is known from, e.g. previous measurements on the cavities. Any unknown variations in this parameter cannot be rectified. 
In the opposite limit of
poor time resolution detectors ($\gamma_r \ll \Delta$) the fidelity
of the resulting mixed entangled state depends on the overlap of the
photons' spectra according to Eq.~(\ref{eqn:Fidt1t2ExplicitExpression}).

With the aid of a model of the internal dynamics of the
photodetector, we were also able to analyse the intermediate case,
where the detector response rate, $\gamma_r$ is of similar order to
$\Delta$. Here, we found that as $\gamma_r$ exceeds $\Delta$, the
fidelity of the entangled states approaches unity. The numerical
results suggest that the error in the entangling operation (that is,
$1-\bar{F}$, with $\bar{F}$ the average fidelity of
Eq.~(\ref{eqn:FidAv2})) scales approximately quadratically with
$\Delta/\gamma_r$, such that an error of $1-\bar{F} \sim 10^{-4}$ is
possible with $\Delta/\gamma_r \sim 10^{-2}$.

We also analysed the Hong-Ou-Mandel effect with detuned photons, and
arrived at similar conclusions. In the case of idealized detectors
with time resolution much larger than the photon detuning, a
modified Hong-Ou-Mandel effect can be observed, with the coalescence
probability oscillating as a function of time, at angular frequency
given by $\Delta$, the detuning of the two sources. This result is
in agreement with earlier theoretical and experimental results by
Legero {\em et al.} \cite{Legero2003,Legero2004}, although our approach was slightly different to this earlier work. We also
extended the analysis to the case of non-ideal detectors, and found
that the fringe visibility of this modified Hong-Ou-Mandel effect
decreases dramatically as $\gamma_r$ is reduced below $\Delta$, but
approaches unity for $\gamma_r \gg \Delta$.

The usual understanding of few-photon interference effects
is that they can only be observed with identical photons. This is
because experiments such as those considered here make use of a
beam splitter to coherently erase `which path' information. Thus any
`excess' information, such as frequency or polarization,
carried by the photon, that can be used, even if only in principle,
to infer the previous path of the photon, will typically degrade the
interference effect. Even if the detector has no output
corresponding to the energy of the incident photon, we could imagine
that this information might be encoded in the internal state of the
detector, e.g.~in the energy of the exciton-hole pair created when
the photon was initially absorbed, and so it could be determined, in
principle, with a careful measurement of the microscopic state of
the detector.

This is in apparent contradiction with the results of this paper,
which predict that a modified version of such interference effects
can be observed with high fidelity even when the photons are
spectrally distinct. How can the results here be reconciled with the
conventional understanding described in the previous paragraph? One
way of understanding this is that detectors with sufficiently good
time resolution are themselves capable of erasing some of this
`excess' frequency information. Loosely speaking, for a detector
with time resolution $\gamma_r^{-1}$, a detected photon is localized
in a time window of corresponding duration $\gamma_r^{-1}$. Thus the
frequency uncertainty of the detected photon \emph{must} be equal to or larger
than $\gamma_r$. This means that there is no way, even in principle,
of determining the frequency of the detected photon. Thus the
`excess' frequency information has been successfully erased.

Our results should also shed light on the question of where the
interference `goes' in few photon interference experiments with
detuned photons. With idealized detectors with high temporal
resolution, we have seen that a modified version of the interference
or entanglement is present, and in this sense there is no
loss of interference visibility or entanglement. With imperfect
detectors, the corresponding effects are diminished. Therefore another way
of understanding these results is to say that imperfect detectors
introduce noise into the interference pattern, because the timing of
the detector clicks produces additional randomness above what would be
expected from spontaneous emission alone.

We conclude by suggesting some possibilities for future work on the
subject of few photon interference with detuned photons. One
immediate application of our work is that it can be used to place
design constraints on near future experiments - given detectors with
a particular temporal resolution (as quantified by $\gamma_r$ in
this work), we know that the sources must be tuned such that $\Delta
\lesssim \gamma_r$ in order to see the corresponding interference or
entanglement effect with reasonable visibility/fidelity. Conversely,
if tuning the sources is difficult, we know the corresponding
detector resolution required.

Similarly, for scalable quantum
computation, the results here can be used to place constraints that
the physical parameters $\Delta$ and $\gamma_r$ must satisfy in
order to implement a fault-tolerant computation scheme, either in a
hybrid matter-light setup \cite{BarrettKok2005}, or in a linear
optical scheme \cite{KLM,KokReview,BrowneRudolph2005}. In this
case, some further work is required in order to relate the physical
errors due to frequency mismatch and finite detector response rate
to the more abstract error models employed in the analysis of fault
tolerant computation schemes.

Finally, although we have concentrated on two particular examples of
few photon interference in this paper, the results presented here,
the methods used to arrive at them, and the resulting physical
insights, should carry over to many other few photon interference
setups. We therefore hope these results have useful implications for
quantum information processing schemes which utilize few photon
interference.

{\em Acknowledgments}:  We thank Jeremy O'Brien, Andrew Doherty, Peter Knight and
in particular Tom Stace for a number of stimulating and encouraging
conversations. SDB was supported by the {\sc epsrc}. JM was
supported by the {\sc epsrc} through the Quantum Information
Processing IRC.

\bibliography{FrequencyMismatch}

\end{document}